\documentclass[conference]{IEEEtran}
\IEEEoverridecommandlockouts
\usepackage{cite}
\usepackage{amsmath,amssymb,amsfonts}
\usepackage{algorithmic}
\usepackage{graphicx}
\usepackage{textcomp}
\usepackage[table]{xcolor}
\usepackage[bottom]{footmisc}
\begin{document}

\title{Diversity in Network-Friendly Recommendations\\}

\author{\IEEEauthorblockN{Evangelia Tzimpimpaki}
\IEEEauthorblockA{\textit{Technical University of Crete} \\
Chania, Greece \\
etzimpimpaki@tuc.gr}
\and
\IEEEauthorblockN{Thrasyvoulos Spyropoulos}
\IEEEauthorblockA{\textit{Technical University of Crete} \\
Chania, Greece \\
spyropoulos@tuc.gr}
}

\maketitle

\IEEEpubid{\begin{minipage}[t]{\columnwidth}
\vspace{0.5cm}
\footnotesize
The authors have been supported in part by the Hellenic
Foundation for Research \& Innovation (HFRI) project 8017:``AI4RecNets:
Artificial Intelligence (AI) Driven Co-design of Recommendation and Networking
Algorithms”
\end{minipage}}

\begin{abstract}
In recent years, the Internet has been dominated by content-rich platforms, employing recommendation systems to provide users with more appealing content (e.g., videos in YouTube, movies in Netflix). While traditional content recommendations are oblivious to network conditions, the paradigm of Network-Friendly Recommendations (NFR) has recently emerged, favoring content that improves network performance (e.g. cached near the user), while still being appealing to the user. However, NFR algorithms sometimes achieve their goal by shrinking the pool of content recommended to users. The undesirable side-effect is reduced content diversity, a phenomenon known as ``content/filter bubble'',\cite{r1,r2}. This reduced diversity is problematic for both users, who are prevented from exploring a broader range of content, and content creators (e.g. YouTubers) whose content may be recommended less frequently, leading to perceived unfairness. In this paper, we first investigate - using real data and state-of-the-art NFR schemes - the extent of this phenomenon. We then formulate a ``Diverse-NFR'' optimization problem (i.e., network-friendly recommendations with - sufficient - content diversity), and through a series of transformation steps, we manage to reduce it to a linear program that can be solved fast and optimally. Our findings show that Diverse-NFR can achieve high network gains (comparable to non-diverse NFR) while maintaining diversity constraints. To our best knowledge, this is the first work that incorporates diversity issues into network-friendly recommendation algorithms.
\end{abstract}

\section{INTRODUCTION}
\noindent {\bf Background.} The majority of content services (video/movie platforms, social networks, e-commerce, etc.) integrate recommendation systems (RS) that significantly influence users' choices among the vast array of available content. These systems have been demonstrated to shape content demand, with 50\% of YouTube's views\cite{r3}, 80\% of Netflix's views\cite{r4}, and 35\% of Amazon's revenue\cite{r5}, coming from their RS. However, this content is delivered through a complex network system, where some items are distant from the user and incur high costs to fetch, while other items may be closer and require minimal expense. A prevalent example of this is Content Delivery Networks (CDNs),\cite{e22}, where an item might be either available on a nearby CDN server (thus incurring lower network costs) or it must be retrieved from the origin/deeper servers (which can be substantially more ``costly'' for the network, the user, and/or the content provider). Unfortunately, standard recommendation systems, referred to hereafter as ``Baseline RS (BSR)'', do not consider this network cost and recommend content solely based on user-centric metrics (e.g. personalization and relevance,\cite{r2}) or business metrics (e.g. revenue,\cite{r1}). To address this issue, the paradigm of \emph{Network-Friendly Recommendations} (NFR) has emerged as a solution to enhance the cost-efficiency of content delivery,\cite{r9,e10,e12,e13,e15,e16}. The primary concept behind NFR is to steer recommendations towards content that can be delivered in a ``network-friendly” manner (e.g. cached in the mobile edge,\cite{r9,e10}), thereby influencing user demand in favor of such content. \smallskip

\noindent {\bf The problem: Diversity in NFR.} Favoring network-friendly (e.g. cached) content involves modifying the baseline recommendations of the platform, which could be perceived as quality of experience degradation for one ore more of the parties involved: (i) \emph{users}, might find the modified recommendations as \emph{less relevant/useful} than the original ones\footnote{This is actually not always the case; a less relevant recommendation might be preferred if it offers higher streaming quality\cite{e16}.}; (ii) \emph{content providers} (and their recommendation algorithm team) might perceive such modifications as too intrusive, considerably obfuscating the original goal of the platform RS; (iii) \emph{user and content creators} might protest if only a small subset of items is consistently presented to the pool of users (e.g., mostly cached content to increase cache hit rate). 

Concern (i) captures the recommendation relevance degradation \emph{for a single user}, and countermeasures have been included in most NFR schemes since their inception,\cite{r9,e16}. Concern (ii) relates to a \emph{global} degradation of recommendations to all users, and has been first considered in the recent work of\cite{e17}, where the authors attempt to obtain NFR gains \emph{while minimizing the intrusiveness of the algorithm} (e.g. by putting a hard constraint on the distance between the BSR and the NFR recommendation vectors). Concern (iii), on the other hand, has neither been explicitly measured yet nor addressed. It is not as immediately clear (as (i) and (ii)) that an NFR algorithm does indeed create such ``content bubbles''. Our conjecture is that, to reduce network cost, an NFR algorithm will replace some recommendation probability mass from items outside the cache, with items inside the cache, \emph{at every opportunity where sufficiently relevant alternatives (e.g. satisfying (i)) exist}. By doing this for most users and most items, the majority of the recommendation probability mass will become concentrated around a much smaller pool of content, signaling decreased diversity in the RS. Of course, excessive diversity is not the goal, since it can \emph{conflict with the accuracy} that RSs are designed to achieve. However, \emph{diversity levels/constraints should be adjustable} based on the preferences of the users/content creators—a feature not yet incorporated in NFR schemes. To this end, our goal in this paper is to \emph{explicitly address the issue of diversity in NFR} for both user and content creator satisfaction\footnote{In fact, diverse recommendations can \emph{also} help address business and societal issues by promoting a broader range of content.}.

While one might argue that metrics targeting concerns (i) or (ii) also resolve the issue of diversity, this is not the case. Consider a simplified example with three users and their relevance scores for three items -each by a different creator: user A: \{1,0.8,0.2\}, user B: \{0.8,1,0.9\}, and user C: \{0.3,0.9,1\}. In the standard BSR, item 1 would be recommended to user A, item 2 to user B, and item 3 to user C, maximizing their satisfaction. However, if only item 2 is cached, the NFR would recommend item 2 to all three users to increase cache hit rate. Although relevance scores (0.8, 1, and 0.9 for users A, B, and C) remain high, diversity is reduced threefold as the RS presents only one item instead of three, leading to dissatisfaction among content producers of items 1 and 3. Hence, the metrics addressing concern (i) do not suffice to fix the problem. In a different scenario, after viewing item 1, the BSR suggests items \{7,4,3\} to users 1-3, while the NFR provides cached items \{2,6,5\}. Though both systems are 100\% \emph{diverse} (i.e., 3 different items across 3 recommendations), none of the NFR items align with the BSR, rendering the NFR algorithm ``too intrusive''. Thus, concerns (ii) and (iii) are also clearly distinct. \smallskip

\noindent {\bf Contributions.} Our contributions in this paper include:
\begin{itemize}
    \item {\bf Diversity characterization.} We propose the use of \emph{entropy} as a measure to quantify diversity in recommendations (Section II), and use it to analyze the outcome of various NFR algorithms on real datasets (Section III).
        
    \item {\bf Optimal Diverse-NFR.} We formulate the problem of Diverse-NFR, transform it first to an (equivalent) convex optimization problem (thus guaranteeing that it can be solved optimally), then further transform it to a linear program to further facilitate solution speed (Section IV).
    
    \item {\bf Diversity and Network-Friendliness.} We prove that high network gains can be achieved while maintaining diversity levels comparable to BSR, e.g. reducing the original BSR cost by $10\times$ while only reducing the original BSR diversity by 40\% (Section V).

    \item {\bf Diversity and Fairness.} We prove that (i) compared to Fair-NFR\cite{e17}, our algorithm achieves superior cost-diversity trade-offs by explicitly prioritizing diversity, and (ii) it is possible to address distinct aspects of fairness alongside ``diversity fairness'' by integrating other fairness metrics  (as\cite{e17}) in our Diverse-NFR (Section VI).\smallskip
\end{itemize}

\section{PROBLEM SETUP}
\subsection{Network-Friendly Recommendations}
We examine a content service that incorporates a RS within
its platform. When a user engages (e.g. watches, listens) with content, a list of $N$ recommendations (related to the item just viewed) is provided by the RS, suggesting content to consume next. Below, we outline the general setup used in NFR, \emph{as proposed in}\cite{e16}, with key notations summarized in Table ~\ref{tab1}.\begin{table}[b]
\vspace{-1em}
\caption{Important Notation}
\vspace{-1em}
\begin{center}
\begin{tabular}{cc}
\hline
$\mathcal{K}$  & $\text{ content catalog } ( \rvert \mathcal{K} \rvert = K)$ \\
$\mathcal{C}$   & $\text{ set of cached contents }(\rvert\mathcal{C}\rvert=C)$\\
$N$  & $\text { number of recommendations }$ \\
$u_{ij}$ \,      & $\text{ relevance score for items } i,j $ \\
$r_{ij}^{ALG}$ \,      & $\text{ prob. to recommend item } j \text{ after } i$ \\
$\mathbf{R^{ALG}} $     & $\text{ recommendation matrix containing } r_{ij}^{ALG}$ \\
$c_i$ \,         & $\text{ access cost for item } i, c_i \in \{0,1\}$ \\
$a$ \,         & $\text{ prob. a user follows recommendations } $\\
$q_i^{max}$ \,         & $\text{ max BSR quality for item }, i$ \\
$q $\,         & $\text{ percentage of } q_i^{max} \text{ offered by NFR }$\\  
top-N \, & the N items with the highest scores $u_{ij}$\\
$\mathbf{p_0}$      & $\text{ direct demand (also, initial total demand)}$ \\
\emph{pop} \, & the Zipf parameter of $\mathbf{p_0}$\\
$\mathbf{p^{ALG}}$      & $\text{ total content demand (ALG is: BSR/NFR)}$ \\
\hline
\end{tabular}
\label{tab1}
\end{center}
\end{table}
{\bf Content Service.} A user requests content from a catalogue \( \mathcal{K} \): (i) by following one of the \( N \) recommendations provided by the RS, with probability \( \alpha \in (0,1) \), or (ii) directly, e.g., by searching through a search bar, with probability $1 - \alpha$. In the first case, he selects among the \( N \) recommendations uniformly\footnote{A generalization of this model, where the position a recommendation is placed in changes the click probability, has been proposed and solved in\cite{e12}. Our diverse-NFR approach is applicable to that model as well, but we focus here on the simpler uniform case to better elucidate our contribution.}. In the second case, the probability he requests an item \( j \) is \( p_{0j} \in (0,1) \); hence, \( \mathbf{p_0} = [p_{01}, \ldots, p_{0K}] \) represents the probability distribution for direct content demand. We assume that \( \mathbf{p_0} \) follows a Zipf distribution, with a Zipf parameter denoted as \emph{pop}, and that it also governs the first content accessed in the session. \\
{\bf Content Demand.} Similarly with $\mathbf{p_0}$, the probability distribution for the \emph{total content demand} (i.e., both direct and through recommendations) $\mathbf{p^{ALG}}$ is calculated for every item, based on the user's choices and the algorithm (BSR/NFR) used by the RS. We elaborate in Section IV how this is derived. For now, we can consider this as a standard measure of ``how popular'' is each content in the specific algorithm. \\
{\bf Network.} Every item is associated with some cost $c_i \in [0,1]$ that depends on where the content is cached (the closer to the user the smaller $c_i$).\\
{\bf Recommendations.} We assume that when a user watches a content $i$, then a list of N items is recommended. Specifically, a vector of $r_{ij}^{ALG}\in [0,1]$ is chosen, such that $\sum_j r_{ij}^{ALG} = N$. We also define the matrix $\mathbf{R^{ALG}}$ containing these recommendation probability vectors for each $i\in \mathcal{K}$.\\ 
{\bf Baseline RS (BSR).} The standard RS that generates the relevance scores $u_{ij}\in [0,1]$, indicating the suitability of recommending content j after content i; higher $u_{ij}$ values represent better recommendations. After a user consumes item $i$, the BSR recommends the top-N items achieving maximum quality of recommendations $q_i^{max} = \sum_{j=1}^K r_{ij}^{BSR} u_{ij}$. Hence, for the highest $u_{ij}$ values, $r_{ij}^{BSR} = 1$, else $r_{ij}^{BSR} = 0$.\\
{\bf Network-Friendly RS (NFR).} NFR (i) includes ``cheaper'' content (i.e. the lowest $c_i$ possible), and (ii)  attempts to (still) recommend ``good enough'' content (according to the underlying platform's $u_{ij}$ scores and its BSR quality $q_i^{max}$)\footnote{It is important to stress that, almost all existing NFR algorithms (including ours), use these $u_{ij}$ values (possibly user-specific) as input to a larger optimization problem, and do not attempt to modify or ``improve'' the workings of the underlying platform algorithm.}.\smallskip

\subsection{Diversity Definition}
\noindent {\bf Diversity.} In order to both measure (Section III) and manipulate (Section IV) diversity, we need to define an appropriate metric. Given that recommendations in both NFR and BSR can be interpreted as probabilities, a natural measure of diversity is entropy. Higher entropy values imply avoiding deterministic patterns where only certain contents are recommended; hence, \emph{diverse recommendations}. Thus, we aim to maintain a relatively high level of entropy\footnote{While one could argue that a goal could be to maximize entropy, hence diversity, this is seldom the (main) goal of a recommender. Instead, we will use the entropy of the baseline scheme as the ``ideal'' diversity one wants to maintain, while reducing network cost.}. To this end, we can calculate the entropy based on: the recommendation probability of content \( i \) following content \( j \) \( \left( r_{ij}^{ALG} \right) \):
\begin{equation}
    H(\mathbf{R^{ALG}}) := - \sum_{j=1}^K r_{ij}^{ALG} \cdot \log(r_{ij}^{ALG}), \forall i \in \mathcal{K}  \label{eq:1} \tag{1}
\end{equation}
or, the long-term probability that \( i\) is requested \( \left( p_i^{ALG}\right) \):
\begin{equation}
    H(\mathbf{p^{ALG}}) := - \sum_{i=1}^K p_i^{ALG} \cdot \log(p_i^{ALG})  \label{eq:2} \tag{2}
\end{equation}

The interdependence between the content demand distribution \( \mathbf{p^{ALG}} \) and recommendation matrix \( \mathbf{R^{ALG}} \)—where enforcing diversity in \( \mathbf{R^{ALG}} \) influences content demand and vice versa—enables both interpretations. Although the first is more intuitive, we will adopt the second interpretation in Section IV for implementation convenience.\smallskip

\section{DECREASED DIVERSITY IN NFR}
Our first goal is to investigate whether, and to what extent, content diversity is indeed affected by making recommendations network-friendly. To this end, we will apply the BSR and NFR algorithms of\cite{e16}, on the Last.FM\cite{e19} and MovieLens\cite{e20} datasets. The parameters of the (total 144) considered scenarios are summarized in Table \ref{tab2}.

\subsection{Simulations Setup}
\noindent {\bf Content catalogs.} We consider content catalogs from the Last.fm ($K = 757$)\cite{e19} and MovieLens ($K = 1060$)\cite{e20} platforms, with matrices $\mathbf{U}$ extracted from them and pre-processed according to\cite{e17}, so that all values $u_{ij} < 0.5$ are set to zero (i.e. irrelevant contents).\\
{\bf Session Length.} We assume an average user session length equal to $L = 40$ items.\\
{\bf Recommendations.} We consider two example values for $N$, the number of recommendation, $N=2$ (e.g., small mobile screen) and $N=10$ (e.g., laptop device).\\
{\bf Caching.} We consider cache sizes $C = \{5,20\}$, storing the items with the highest BSR demand (i.e., the highest $\mathbf{p^{BS}}$). \\
{\bf Content Service.} We consider $\alpha = \{0.5, 0.8,0.99\}$, to capture the behavior for YouTube\cite{r3}, Netflix\cite{r4}, and an extreme case where users almost always follow recommendations (e.g. autoplay) respectively. We also run simulations for two direct-demand Zipf parameter values: \emph{pop}$ =0$ (equal probability for all items) and \emph{pop}$ =1$ (higher probability for the first items). \\
{\bf QoR.} We consider quality percentages $q = \{ 0.5, 0.8, 0.99\}$.\smallskip

\begin{table}[t] 
\vspace{-1em} \caption{Simulation parameters} \vspace{-1em}
	\begin{center}
	\begin{tabular}{c}
	\hline $\mathbf{U}:  \text { Last.Fm } (K=757) \text {, MovieLens } (K=1060)$ \\
	$L = 40, \quad \, N \in \{2,10\}, \quad  C \in \{5,20\} $ \\
	$\alpha \in \{0.5, 0.8, 0.99\}, \quad \, \textit{pop} \in \{0,1\}, \, q = \{ 0.5, 0.8, 0.99\}$\\  
	\hline \vspace{-2em} 
	\end{tabular} \label{tab2}
	\end{center}
\end{table} \smallskip

\subsection{Decreased diversity in simulations}
In the scenarios we simulate, we calculate the content demand under the BSR ($\mathbf{p^{BS}}$) and the NFR algorithm ($\mathbf{p^{NF}}$), and subsequently compute the entropy of each system, as defined in Eq. (2) of Section II.B.\smallskip

Table \ref{tab3} presents the entropy and network cost values from the first simulations. Each pair includes data for the BSR and the NFR of the same input parameters (specified in the second column), with a double line separating the distinct pairs. The third column displays the network cost for the respective RS and the fourth column expresses this cost as a percentage of the BSR network cost. The entropy of the RS is computed in the fifth column, followed by the percentage this entropy holds on the BSR entropy.\smallskip
\begin{table}[b]
\caption{NFR entropy $(H)$ and network cost $(c)$ as percentages of the respective BSR values, in 5 different scenarios.} \vspace{-2em}
\begin{center}
\begin{tabular}{|c|c|c|c|c|c|}
\hline 
\rule[-1ex]{0pt}{2.5ex} {RS} & {Parameters} & {$c$} & {\%$c^{BS}$} & {$H$} & {\%$H^{BS}$} \\
\hline  
\rule[-1ex]{0pt}{2.5ex} {BS} & {pop0 a.99 N2 C20 q0.8} & {0.817} & {100\%} & {5.808} & {100\%}\\ 
\hline 
\rule[-1ex]{0pt}{2.5ex} {NF} & {pop0 a.99 N2 C20 q0.8} & {0.032} & {3.9\%} & {1.938} & {\bf 33\%}\\ 
\hline \hline 
\rule[-1ex]{0pt}{2.5ex} {BS} & {pop0 a0.99 N10 C20 q0.8} & {0.687} & {100\%} & {5.628} & {100\%}\\ 
\hline 
\rule[-1ex]{0pt}{2.5ex} {NF} & {pop0 a0.99 N10 C20 q0.8} & {0.337} & {49\%} & {3.829} & {\bf 67\%}\\ 
\hline \hline 
\rule[-1ex]{0pt}{2.5ex} {BS} & {pop0 a0.80 N2 C20 q0.5} & {0.891} & {100\%} & {6.312} & {100\%}\\ 
\hline 
\rule[-1ex]{0pt}{2.5ex} {NF} & {pop0 a0.80 N2 C20 q0.5} & {0.271} & {30\%} & {4.070} & {\bf 64\%}\\  
\hline \hline 
\rule[-1ex]{0pt}{2.5ex} {BS} & {pop1 a0.80 N2 C20 q0.8} & {0.790} & {100\%} & {5.917} & {100\%}\\ 
\hline 
\rule[-1ex]{0pt}{2.5ex} {NF} & {pop1 a0.80 N2 C20 q0.8} & {0.329} & {41\%} & {4.130} & {\bf 69\%}\\  
\hline \hline 
\rule[-1ex]{0pt}{2.5ex} {BS} & {pop1 a0.99 N2 C20 q0.8} & {0.787} & {100\%} & {5.745} & {100\%}\\ 
\hline 
\rule[-1ex]{0pt}{2.5ex} {NF} & {pop1 a0.99 N2 C20 q0.8} & {0.033} & {4\%} & {2.349} & {\bf 40\%}\\  
\hline
\end{tabular}
\label{tab3}
\end{center}
\end{table}

\begin{figure}[t]
\centerline{\includegraphics[scale=0.6]{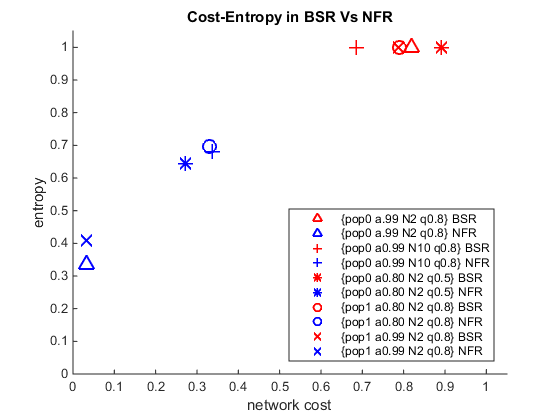}}
\caption{Normalized Network cost (x-axis) vs. Entropy (y-axis) in NFR (blue points) and BSR (red points) across five Last.FM scenarios with $C = 20$. All cost and entropy values
have been normalized as a percentage of the respective BSR values.} \vspace{-1em}
\label{fig1}
\end{figure}

Fig.\ref{fig1} illustrates the entropy and network cost for both BSR and NFR across various parameters (e.g., for the circular markers, the parameters are: \{\emph{Last.FM, pop1, a0.8, N2, C20, q0.8}\}). Red markers represent the BSR values, while blue markers indicate the corresponding NFR values for the same parameters. We observe that: (i) the BSR points are concentrated in the upper-right quadrant, where both costs and entropy are high, and (ii) the NFR points shift towards the lower-left quadrant, reflecting reduced costs but also decreased diversity. \smallskip

\noindent The key findings of these initial results can be summarized as:\\

\noindent {\bf Key Finding 1: \textit{Shifting from BSR to NFR - for network efficiency - consistently results in reduced content diversity across all the scenarios we examined.}}\\

\noindent {\bf Key Finding 2:} {\bf \textit{The available trade-offs seem pessimistic: scenarios with higher cost savings often result in the greatest diversity losses}}. The extent of this trade-off varies by scenario and is influenced by system parameters such as the number of recommendations (\(N\)), cache size (\(C\)), and user behavior (\(\alpha\)), which are set for each simulation. For example, in the first case of Table \ref{tab3}, entropy is reduced to 33\% of the BSR entropy. However, in other cases, such as the second scenario of Table \ref{tab3}, entropy remains relatively high at 67\% of the BSR entropy. This outcome can be explained by considering the actual meaning of these parameters.\smallskip

In the first scenario, a high \(\alpha\) value represents a user ``flexible'' in content consumption, i.e., consistently following recommendations. This scenario allows for significant network cost reductions. In order to achieve this cost efficiency, the RS has very few low-cost content options available. Therefore, it consistently recommends items from this limited set, and diversity in recommendations is diminished. \smallskip

On the other hand, in the second scenario, the large number of recommendations (\(N\)) makes it difficult to find enough cached items to recommend even though the user is still ``flexible'' in content consumption. As a result, the system must recommend non-cached content, thereby increasing network costs. Since the cost is already relatively high and cannot be further minimized, there is a broader set of items available to satisfy the cost constraints, allowing for more diverse recommendations.\smallskip

\section{OPTIMAL DIVERSE NFR}

In this section, we revisit the fundamental NFR Optimization Problem (OP) of\cite{e16}, with a new objective: achieve high diversity while still reducing network costs. To this end, our initial Diverse-NFR OP is (3)\footnote{We use $\mathbf{R}$, $r_{ij}$ as an abbreviation of $\mathbf{R^{NFR}}$, $r_{ij}^{NFR}$ respectively.}. The black part of (3) is the original NFR OP formulation of\cite{e16}, and the requirement for diversity is captured by introducing a new constraint (3.e), shown in red color, that will ensure that our Entropy-based diversity metric, (1), of Section II.B remains above a required percentage, b, of the original (BSR) entropy\footnote{One could also include this in the objective as a ``soft constraint'' but the convexity of the problem should still be proven.}. \smallskip

The NFR decides which contents to recommend by using the set of variables $r_{ij}$, representing the probability that content $j$ will be recommended after content $i$, as proposed in\cite{e16}. By using a probabilistic approach, where $r_{ij} \in [0,1]$, the algorithm can vary the recommendations between different users or even for the same user over time, helping to avoid repetitive suggestions for a particular content.\smallskip

The objective of the NFR is the minimization of the expected cost for a long user session, calculated as (3.a). The objective calculation was done in\cite{e16}, where a user session is modeled as an absorbing Markov chain. Each session consists of several recommendation-following periods ($S_R$), which can end at any step with a probability of \(1 - a\) (the absorbing state). After a recommendation period concludes, the process ``renews," with the user re-entering the catalog from the same initial distribution \(\mathbf{p_0}\), making each $S_R$ period i.i.d. \smallskip

Last, the NFR constraints include that: a certain quality of recommendations (QoR) is achieved (3.b), each recommendation list contains exactly $N$ items (3.c), and the decision variables are valid probabilities (3.d). \\

\noindent{\bf Diverse-NFR OP} 
\begin{align}
& \underset{\mathbf{R}}{\operatorname{minimize}} \quad  \frac{\mathbf{p}_0^T \cdot\left(\mathbf{I}-\frac{\alpha}{N} \cdot \mathbf{R}\right)^{-1} \cdot \mathbf{c}}{\frac{1}{1-\alpha}}  \label{eq:3.a} \tag{3.a} \\
& \text{subject to} \quad \sum_{j=1}^K r_{i j} \cdot u_{i j} \geq q \cdot q_i^{max}, \quad \forall i \in \mathcal{K} \label{eq:3.b} \tag{3.b} \\
& \qquad \qquad \quad\sum_{j=1}^K r_{i j}=N, \quad \forall i \in \mathcal{K} \label{eq:3.c} \tag{3.c} \\
& \qquad \qquad \quad 0 \leq r_{i j} \leq 1 \, (i \neq j), \quad r_{i i}=0 \label{eq:3.d} \tag{3.d} \\
& \qquad \qquad \quad {\color{red} -\sum_{j=1}^K r_{ij} \cdot \log(r_{ij}) \geq b \cdot H(\mathbf{R^{BS}}) , \, \forall i \in \mathcal{K} \label{eq:3.e} \tag{3.e}}
\end{align}

Any new constraint introduced into an optimization problem might (considerably) reduce the feasible region for the recommendation variables (and thus the network cost reduction achievable), hence needs to be further explored. What is worse, Problem 3 above is not even convex with respect to $\mathbf{R}$, due to the inverse matrix in the objective function. The authors of\cite{e16} have introduced a change of variables to convexify the problem, but it is not clear whether this change of variables will  maintain the convexity of the new constraint (3.e) or not. To this end, we apply the variable transformation proposed in\cite{e16} which included two parts: (i) introducing the NFR long term demand $\mathbf{p^{NF}} = (1-\alpha) \cdot \mathbf{p}_0^T \cdot \left(\mathbf{I}-\frac{\alpha}{N} \cdot \mathbf{R}\right)^{-1}$ as an auxiliary optimization variable, and (ii) applying the transformation $f_{ij} = r_{ij} \cdot p^{NF}_i$ to convert the new, non-convex constraint which arises from (i):
\begin{align*}
p^{NF}_j = (1 - \alpha) \cdot p_{0j} + \frac{a}{N} \sum_{i = 1}^K r_{ij} \cdot p^{NF}_i 
\end{align*}

to a convex one: 
\begin{align*}
 p^{NF}_j = (1 - \alpha) \cdot p_{0j} + \frac{a}{N} \sum_{i = 1}^K f_{ij}.
\end{align*}

Then, we examine the convexity of the new OP which includes our diversity constraint. Hence, our first result is the proof of the following lemma.\\

\subsection{Convex Diverse-NFR: Proof of convexity}

\noindent {\bf Lemma 1.} \emph{The following equivalent OP, (4), is convex with respect to the new optimization variables $\mathbf{F}, \mathbf{p^{NF}}$.}\\

\noindent{\bf Convex Diverse-NFR OP} 
\begin{align}
& \underset{\mathbf{p^{NF}, \mathbf{~F}}}{\operatorname{minimize}} \quad \mathbf{c}^T \mathbf{p^{NF}} \label{eq:4.a} \tag{4.a} \\
& \text{subject to} \quad \sum_{j=1}^K f_{i j} \cdot u_{i j} - p_i^{NF} \cdot q \cdot q_i^{max} \geq 0, \forall i \in \mathcal{K} \label{eq:4.b} \tag{4.b} \\
& \qquad \qquad \quad \sum_{j=1}^K f_{i j} - N \cdot p_i^{NF} = 0, \forall i \in \mathcal{K} \label{eq:4.c} \tag{4.c} \\
& \qquad \qquad \quad f_{i j} - p_i^{NF} \leq 0, \forall i, j \in \mathcal{K} \label{eq:4.d} \tag{4.d} \\
& \qquad \qquad \quad f_{i j} \geq 0 \, (i \neq j), \, f_{i i}=0 \label{eq:4.e} \tag{4.e} \\
& \qquad \qquad \quad p_j^{NF} - \frac{\alpha}{N} \cdot \sum_{i=1}^K f_{i j} = p_{0 j}, \forall j \in \mathcal{K} \label{eq:4.f} \tag{4.f} \\
& \qquad \qquad \, {\color{red} - \sum_{j=1}^K f_{ij} \cdot log(\frac{f_{ij}}{p^{NF}_i}) \geq b \cdot  H(\mathbf{R^{BS}}) \cdot p^{NF}_i, \forall i \in \mathcal{K}} \label{eq:4.g} \tag{4.g}
\end{align}

\noindent{\bf \emph{Proof.}} The convexity of the objective and the constraints (4.b)-(4.f) is straightforward, and proven more formally in\cite{e16}. We will focus on proving the convexity for our diversity constraint, (4.g). To this end, we first define the function: $$ \phi_i(f_{ij},p^{NF}_i) = \sum_{j=1}^K f_{i j} \cdot log(\frac{f_{ij}}{p^{NF}_i}) + b \cdot H(\mathbf{R^{BS}}) \cdot p^{NF}_i.$$ To ensure that adding (4.g) to the OP maintains the problem's convexity, we need to prove that \( \phi_i \) is \emph{convex} within its domain. To establish the convexity of \( \phi_i\), we rely on the principle that the sum of convex functions is itself convex. Thus, we focus on proving that \( k_i(f_{ij}, p^{NF}_i) = f_{ij} \cdot \log\left(\frac{f_{ij}}{p^{NF}_i}\right)\) is convex. This can be done by applying the Second-Order Condition of Convexity for \( k_i \), at all points \( f_{ij}, p^{NF}_i \in \text{dom}k_i \), where $domk_i = \{ f_{ij} > 0, p_i^{NF} > 0, \forall i,j \in \mathcal{K}\}$:\\

\noindent \textit{Second-Order Condition of Convexity}: A twice continuously differentiable function $f$, with an open convex
domain $dom f$, is convex if and only if $\nabla^2 f(\mathbf{x}) \succcurlyeq \mathbf{0}, \forall \mathbf{x} \in domf$.  \\

So, we calculate the gradient and the Hessian of $k_i$ as\footnote{In the calculations of the gradient and the Hessian of $k_i$, we have replaced $p^{NF}_i$ with the shorter $z_i$, and we shall do it for the rest of the proof for briefness.}
:\\

\small \noindent$
\nabla k_i(f_{ij},z_i) 
= \left[\begin{array}{c}
\dfrac{d k_i(f_{ij},z_i)}{d f_{ij}} \\
\, \\
\dfrac{d k_i(f_{ij},z_i)}{d z_i} \\
\end{array}\right] 
= \left[\begin{array}{c}
logf_{ij} + 1 - logz_i \\
\, \\
-  \dfrac{f_{ij}}{z_i} \\
\end{array}\right]
$

\normalsize \, \\

\small \noindent $\nabla^{2} k_i(f_{ij},z_i) 
=
\left[\begin{array}{cc}
  \dfrac{d^{2} k_i}{d f_{ij}^2}
& \dfrac{d^{2} k_i}{d f_{ij} \, \, d z_i} \\ \\
  \dfrac{d^{2} k_i}{d z_i \, \, d f_{ij}} 
& \dfrac{d^{2} k_i}{d z_i^2}
\end{array}\right] 
=
\left[\begin{array}{cc}
  \dfrac{1}{f_{ij}}
& \dfrac{-1}{z_i} 
\\ \\
 \dfrac{-1}{z_i} 
& \dfrac{f_{ij}}{z_i^{2}}
\end{array}\right]
$ 

\, \normalsize \\

Having calculated $\mathbf{A} = \nabla^{2} k_i(f_{ij},z_i)$, we prove that it is positive-definite ($\mathbf{A} \succcurlyeq \mathbf{0}$) using Sylvester's criterion.\\

\textit{Sylvester's criterion}: A symmetric Hermitian matrix is positive semi-definite if and only if all its leading principal minors are non-negative.\\

$\mathbf{A}$ \textit{is symmetric} (${a_{ij} = a_{ji}}$), and its principal minors are: 
\begin{itemize}
\item[•] ${\bf A}_{1,1} = 1/f_{ij} > 0$  
\item[•] ${\bf A}_{2,2} = f_{ij}/z_i^2 > 0 $
\item[•] $det({\bf A}) =  \dfrac{1}{f_{ij}} \dfrac{f_{ij}}{z_i^{2}} - \dfrac{1}{z_i} \dfrac{1}{z_i} =  \dfrac{1}{z_i^{2}} - \dfrac{1}{z_i^2} = 0$
\end{itemize} 

\, \smallskip

Since all principal minors of $\mathbf{A}$ are non-negative, it is positive semi-definite, according to Sylvester's criterion. Additionally, $domk_i$ is convex, so $k_i(f_{ij},z_i) = f_{i j} \cdot log(\frac{f_{ij}}{z_i})$ is convex $\forall i,j \in \mathcal{K}$ within its domain, ensuring the overall convexity of (4).\\

The good news is that we now have a convex \emph{equivalent} problem, which is \emph{theoretically solvable}. However, solving large versions of this problem may still be computationally intensive. Thus, we aim to further transform the problem into an LP—since the logarithmic function makes \( \phi_i \) non-linear—so we can leverage well-known LP solvers (specifically, CPLEX) that provide optimality guarantees and faster execution times.\\

\subsection{Linear Approximation of Convex Diverse-NFR}
The first step of our \emph{linear approximation} was inspired by the approach in\cite{e17}, where the authors approximated the logarithmic function with a linear one. However, our problem introduces additional complexity, as we need to linearize a \emph{product containing a logarithmic term}.\\

Similar to \cite{e17}, our first step involved introducing a set of \( K^2\) variables, \( \mathbf{d} \in \mathbb{R}^{K \times K} \), and formulate the following set of inequalities to replace (3.e):\smallskip
\begin{align}
& \sum_{j=1}^K d_{ij} \leq -b \cdot H(\mathbf{R^{BS}}), \, \forall i \in \mathcal{K} \label{eq:5.a} \tag{5.a} \\
& r_{ij} \cdot \log (r_{ij}) \leq d_{ij}, \, \forall i,j \in \mathcal{K} \label{eq:5.b} \tag{5.b} 
\end{align}  \smallskip

Unfortunately, (5.b) is non-linear in \( r_{ij} \). To transform it into a linear form, we approximate the non-linear terms \( r_{ij} \cdot \log (r_{ij}) \) using a general family of linear cuts. Specifically, for each pair \( \{i,j\} \), we define \( M \) lines, \( L(r_{ij}) = a_{m,ij} \cdot r_{ij} + b_{m,ij} \) ($m = 1,...,M$), which are tangent to the function \( g(r_{ij}) = r_{ij} \cdot \log (r_{ij}) \) in \( r_{ij} \in (0,1] \) . The tangent lines \( L(r_{ij}) \) are computed using the First-Order Taylor approximation at selected points of $g$.\\

{\bf Definition.} The First-order Taylor approximation of a (multivariate) function $f(\mathbf{x})$ around a point $\mathbf{x_0}$ is given by: $ f(\mathbf{x}) \approx f(\mathbf{x}_0) + \nabla f(\mathbf{x}_0)^T (\mathbf{x} - \mathbf{x}_0)$.\\

The selection of points for $g$ involved a choice between exponential and linear sampling methods. The tangent lines, $L(r_{ij})$, derived from the First-Order Approximation at these points are as follows:\\
 \begin{itemize}
     \item Exponential sampling: \smallskip
     
     Sample $g$ at points $\left\{e^{-(m-1)s}, -(m-1)s e^{-(m-1)s} ) \right\}$. \\
     
     \small $L(r_{ij}) = g' \left(e^{-(m-1)s}\right) (r_{ij} - e^{-(m-1)s}) + g(e^{-(m-1)s})=$\\
     
     \normalsize \qquad \quad  $ = (1 -(m-1)s) r_{ij} - e^{-(m-1)s} $ \\  
         
     \item Linear sampling:\smallskip
     
     Sample $g$ at points $\left\{ \frac{m}{100}, \frac{m}{100}\log(\frac{m}{100}) \right\}$. \\
     
     \small $L(r_{ij}) = g'\left(\dfrac{m}{100}\right) \left(r_{ij} - \dfrac{m}{100}\right) + g\left(\dfrac{m}{100}\right)=$\\
     
     \normalsize \qquad \quad  $ = \left(1 + \log \left(\frac{m}{100}\right)\right) r_{ij} - \frac{m}{100}$\\  
 \end{itemize}

Through both linear and exponential sampling on $g$, we found that linear sampling provided a better approximation function $L(r_{ij})$ of $g$. Therefore, we use:\smallskip

$$
 \begin{aligned}
& L(r_{ij}) = \left(1 + \log \left(\frac{m}{100}\right)\right) r_{ij} - \frac{m}{100}, \\
\end{aligned}
$$ \smallskip

\noindent which is linear on $r_{ij}$ and can replace $g(r_{ij}) = r_{ij} \cdot \log(r_{ij})$ in (5.b). Further applying $r_{ij} = \frac{f_{ij}}{p_i^{NF}}$, (5.b) becomes:

$$
\begin{aligned}
\left(1 + \log\left(\frac{m}{100}\right)\right) f_{ij} - \frac{m}{100} \cdot p_i^{NF} \leq d_{ij} \cdot p_i^{NF},  \forall i,j,m 
\end{aligned} 
$$\smallskip

The above equation is linear in \( f_{ij} \), but quadratic in \( p_i^{NF} \) and \( d_{ij} \). To address this, we opted to \emph{redefine the diversity constraint} to focus on the resulting content demand distribution \( \mathbf{p^{NF}} \) rather than directly on the recommendation matrix \( \mathbf{R} \), as outlined in Section II.B. This approach allows us to effectively handle the associated linearization challenge. Consequently, equation (3.e) was reformulated as:
\begin{equation} 
 {H(\mathbf{p^{NF}}) = - \sum_{i=1}^K p_i^{NF} \cdot \log(p_i^{NF}) \geq b \cdot H(\mathbf{p^{BS}})} \label{eq:6} \tag{6} \\
\end{equation}
Repeating all the steps of the linear approximation method above for \( g(p_i^{NF}) = p_i^{NF} \cdot \log(p_i^{NF}) \), the linear equivalent of $g$ is $L(p_i^{NF}) = \left(1 + \log\left(\frac{m}{100}\right)\right) p_i^{NF} - \frac{m}{100}$, and (6) becomes:\smallskip
\begin{align} 
& \sum_{i=1}^K d_i \leq -b \cdot H(\mathbf{p^{BS}})  \label{eq:7.a} \tag{7.a} \\
& \left(1 + \log\left(\frac{m}{100}\right)\right) p_i^{NF} - \frac{m}{100} \leq d_i,
 \, \forall i,m \label{eq:7.b} \tag{7.b} 
\end{align}\\
Replacing (4.g) with (7) in the Convex Diverse-NFR OP (4), yields the \emph{\bf Linear Diverse-NFR OP}, (8). This formulation is not an \emph{exact} equivalent of the original problem but rather a linearized approximation. However, it \emph{does} become equivalent in the limit, necessitating that we determine the appropriate trade-off point between diversity and network cost.\\

{\bf Linear Diverse-NFR OP} 
\begin{align}
& \underset{\mathbf{p^{NF}, \mathbf{~F}}}{\operatorname{minimize}} \quad \mathbf{c}^T \mathbf{p^{NF}} \label{eq:8.a} \tag{8.a} \\
& \text{subject to} \, \sum_{j=1}^K f_{i j} \cdot u_{i j} - p_i^{NF} \cdot q \cdot q_i^{BS} \geq 0, \forall i \in \mathcal{K} \label{eq:8.b} \tag{8.b} \\
& \qquad \qquad  \sum_{j=1}^K f_{i j} - N \cdot p_i^{NF} = 0, \forall i \in \mathcal{K} \label{eq:8.c} \tag{8.c} \\
& \qquad \qquad  f_{i j} - p_i^{NF} \leq 0, \forall i, j \in \mathcal{K} \label{eq:8.d} \tag{8.d} \\
& \qquad \qquad  f_{i j} \geq 0 \, (i \neq j), \, f_{i i}=0 \label{eq:8.e} \tag{8.e} \\
& \qquad \qquad p_j^{NF} - \frac{\alpha}{N} \cdot \sum_{i=1}^K f_{i j} = p_{0 j}, \forall j \in \mathcal{K} \label{eq:8.f} \tag{8.f} \\
& \qquad \qquad  \color{red} \sum_{i=1}^K d_i \leq -b \cdot H(\mathbf{p^{BS}}) \label{eq:8.g} \tag{8.g} \\
& \qquad \qquad \color{red} \left(1 + \log\left(\frac{m}{100}\right)\right) p_i^{NF} - \frac{m}{100} \leq d_i,
 \, \forall i,m \label{eq:8.h} \tag{8.h} 
\end{align}

\, \smallskip

\noindent {\bf Remark.} The tuning parameter \( b \) controls the balance between diversity and cost. As \( b \to 0 \), the diversity constraint is effectively removed, focusing entirely on cost minimization. Conversely, as \( b \to 1 \), the RS converges to the BSR, which is conventionally assumed to prioritize diversity.\\

\section{DIVERSITY AND NETWORK-FRIENDLINESS}
\begin{figure}[t]
\centerline{\includegraphics[scale=0.6]{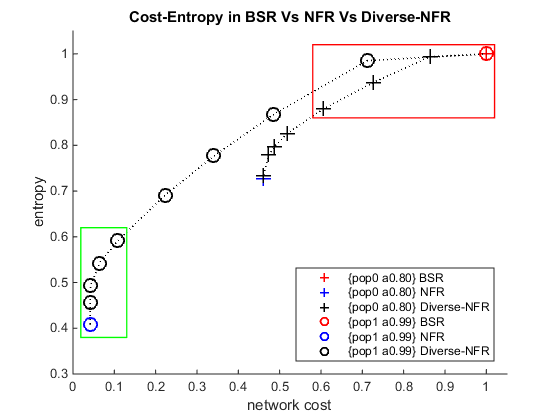}}
\caption{Normalized Network cost (x-axis) Vs Entropy (y-axis) in NFR (blue points), BSR (red points) and Diverse-NFR (black poinst) across two scenarios (denoted with crosses and circles) of the LastFm dataset setups detailed in Section III-A. The values are normalized as a percentage of the respective BSR values.}
\label{fig2}
\end{figure}

In this section, we implement our new Linear Diverse-NFR algorithm, and apply it to the same simulation setup of Section III-A, examining its performance for different values of \( b \) in 2 scenarios\footnote{The simulations included a total of 144 runs with varied simulation parameters, consistently yielding positive results.} (Fig. \ref{fig2}, Table \ref{tab4}, and Table \ref{tab5}). Our key findings are:\\

\noindent {\bf Key Finding 1: \emph{The cost (w.r.t network gain) of imposing diversity constraints is small.}} As network cost increases \emph{slightly}, the entropy initially rises \emph{significantly}, capturing substantial diversity gain with minimal additional cost. This effect is illustrated by the data points in the green square in Fig. \ref{fig2}, with exact percentage values in Table \ref{tab4}. For example, reducing the original BSR cost by 10$\times$  results in only a 40\% reduction in the original BSR diversity (highlighted row in Table \ref{tab4}).\smallskip\\

\begin{table}[b]
\caption{Cost ($c$) and entropy ($H$) values for the specific scenario of \textit{Lastfm pop1 a0.99 N2 C20 q0.8 L40}.}
\begin{center}
\begin{tabular}{|c|c|c|c|c|}
\hline 
\rule[-1ex]{0pt}{2.5ex} {\bf Algorithm} & {\bf c} & {\bf \% of $\mathbf{c^{BS}}$} & {$\mathbf{H}$} & {\bf \% of $\mathbf{H^{BS}}$} \\
\hline 
\rule[-1ex]{0pt}{2.5ex} {\color{blue} NFR} & {\color{blue}0.03275} & {\color{blue}4\%} & {\color{blue}2.34916} & {\color{blue}40\%}\\  
\rule[-1ex]{0pt}{2.5ex} {b = 0.10} & {0.03275} & {4\%} & {2.61673} & {45\%}\\  
\rule[-1ex]{0pt}{2.5ex} {b = 0.50} & {0.03355} & {4.2\%} & {2.83142} & {49\%}\\  
\rule[-1ex]{0pt}{2.5ex} {b = 0.55} & {0.05036} & {6\%} & {3.11591} & {54\%}\\  
\rowcolor[RGB]{193,228,230} \rule[-1ex]{0pt}{2.5ex} {b = 0.60} & {0.08533} & {10\%} & {3.39780} & {60\%}\\  
\rule[-1ex]{0pt}{2.5ex} {b = 0.70} & {0.17530} & {22\%} & {3.96947} & {69\%}\\  
\rule[-1ex]{0pt}{2.5ex} {b = 0.80} & {0.26843} & {34\%} & {4.46336} & {78\%}\\  
\rule[-1ex]{0pt}{2.5ex} {b = 0.90} & {0.38141} & {48\%} & {4.98193} & {87\%}\\  
\rule[-1ex]{0pt}{2.5ex} {b = 1.00} & {0.56048} & {71\%} & {5.65959} & {99\%}\\  
\rule[-1ex]{0pt}{2.5ex} {\color{red} BSR} & {\color{red}0.78742} & {\color{red}100\%} & {\color{red}5.74528} & {\color{red}100\%}\\ 
\hline 
\end{tabular}
\label{tab4}
\end{center}
\end{table}

\noindent  {\bf Key Finding 2: \emph{Substantial network gains can be achieved with a small reduction in diversity.}} Allowing slight relaxations in diversity requirements can result in significant network improvements, as shown by the data points in the red square in Fig. \ref{fig2} , with exact percentage values in Table \ref{tab5}. For instance, for \( b = 1 \) in Table \ref{tab5}, a 0.6\% decrease in diversity leads to a 14\% network gain. Even greater network gains are possible with larger decreases in diversity (for smaller \( b \) values), e.g., a 12\% reduction of BSR diversity allows a 40\% reduction of network costs in the highlighted row of Table \ref{tab5}.\\

\begin{table}[t]
\caption{Cost ($c$) and entropy ($H$) values for the specific scenario of \textit{Lastfm pop0 a0.8 N2 C20 q0.8 L40}.}
\begin{center}
\begin{tabular}{|c|c|c|c|c|} 
\hline 
\rule[-1ex]{0pt}{2.5ex} {\bf Algorithm} & {\bf c} & {\bf \% of $\mathbf{c^{BS}}$} & {$\mathbf{H}$} & {\bf \% of $\mathbf{H^{BS}}$} \\ \hline 
\rule[-1ex]{0pt}{2.5ex} {\color{blue} NFR} & {{\color{blue} 0.41072}} & {\color{blue} 46\%} & {\color{blue} 4.58777} & {\color{blue} 72\%}\\ 
\rule[-1ex]{0pt}{2.5ex} {b = 0.10} & {0.41072} & {46\%} & {4.63642} & {72\%}\\ 
\rule[-1ex]{0pt}{2.5ex} {b = 0.70} & {0.41072} & {46\%} & {4.63682} & {73\%} \\ 
\rule[-1ex]{0pt}{2.5ex} {b = 0.75} & {0.41072} & {46\%} & {4.63750} & {73.5\%} \\ 
\rule[-1ex]{0pt}{2.5ex} {b = 0.80} & {0.42211} & {47\%} & {4.92127} & {78\%} \\ 
\rule[-1ex]{0pt}{2.5ex} {b = 0.82} & {0.43425} & {48\%} & {5.03492} & {79\%} \\ 
\rule[-1ex]{0pt}{2.5ex} {b = 0.85} & {0.46149} & {52\%} & {5.21413} & {82\%} \\ 
\rowcolor[RGB]{253, 200, 252} \rule[-1ex]{0pt}{2.5ex} {b = 0.90} & {0.54023} & {60\%} & {5.55556} & {88\%} \\ 
\rule[-1ex]{0pt}{2.5ex} {b = 0.95} & {0.64721} & {72\%} & {5.91701} & {93\%} \\ 
\rule[-1ex]{0pt}{2.5ex} {b = 1.00} & {0.76969} & {86\%} & {6.27383} & {99.4\%} \\ 
\rule[-1ex]{0pt}{2.5ex} {\color{red} BSR} & {\color{red}0.89078} & {\color{red}100\%} &{{\color{red} 6.31276}} & {\color{red}100\%}\\ 
\hline 
\end{tabular}
\label{tab5}
\end{center}
\end{table}

\noindent {\bf Key Finding 3: \emph{Diverse-NFR achieves a non-linear trade-off between BSR and NFR.}} A concluding observation is that, not only can we explore different intermediate points between the two extremes (red and blue circles), but that there are ``sweet spots'' to explore, where what we lose in one objective does not increase linearly with what we gain in the other, as evidenced by the concave shape of the plots.\\

\section{DIVERSITY AND FAIRNESS}
Having demonstrated that our goal is achievable, we turn our interest in the fairness metrics existing in literature\cite{e17}, and two questions arise that still need to be addressed:\\

\textit{1. Could these other fairness metrics serve the same purpose, thereby rendering our contributions redundant?}\\

\textit{2. Since diversity is not the only form of ``fairness" to consider, would introducing additional fairness goals compromise the favorable trade-offs we've observed so far?}\\

To address these questions, we utilize the fairness metrics from \cite{e17} (summarized in Table \ref{tab6}) and incorporate them into our Linear Diverse-NFR OP. The authors of \cite{e17} have linearized these fairness constraints (as shown in Table \ref{tab7}), enabling their inclusion in our Linear Diverse-NFR OP -without compromising linearity- and resulting in the \emph{Linear Fair-Diverse-NFR OP}, (9). \\ \smallskip

{\bf (Linear) Fair-Diverse-NFR OP} 
\begin{align}
& \underset{\mathbf{p^{NF}, \mathbf{~F}}}{\operatorname{minimize}} \quad \mathbf{c}^T \mathbf{p^{NF}} \label{eq:9.a} \tag{9.a} \\
& \text{subject to} \, \sum_{j=1}^K f_{i j} \cdot u_{i j} - p_i^{NF} \cdot q \cdot q_i^{BS} \geq 0, \forall i \in \mathcal{K} \label{eq:9.b} \tag{9.b} \\
& \qquad \qquad  \sum_{j=1}^K f_{i j} - N \cdot p_i^{NF} = 0, \forall i \in \mathcal{K} \label{eq:9.c} \tag{9.c} \\
& \qquad \qquad  f_{i j} - p_i^{NF} \leq 0, \forall i, j \in \mathcal{K} \label{eq:9.d} \tag{9.d} \\
& \qquad \qquad  f_{i j} \geq 0 \, (i \neq j), \, f_{i i}=0 \label{eq:9.e} \tag{9.e} \\
& \qquad \qquad p_j^{NF} - \frac{\alpha}{N} \cdot \sum_{i=1}^K f_{i j} = p_{0 j}, \forall j \in \mathcal{K} \label{eq:9.f} \tag{9.f} \\
& \qquad \qquad  \color{red} \sum_{i=1}^K d_i \leq -b \cdot H(\mathbf{p^{BS}}) \label{eq:9.g} \tag{9.g} \\
& \qquad \qquad \color{red} \left(1 + \log\left(\frac{m}{100}\right)\right) p_i^{NF} - \frac{m}{100} \leq d_i, \, \forall i,m \label{eq:7.h} \tag{7.h} \\
& \qquad \qquad \, \mathbf{S(z,p^{NF})} \label{eq:9.i} \tag{9.i} 
\end{align}

\, \smallskip

\begin{figure}[t]
\centerline{\includegraphics[scale=0.6]{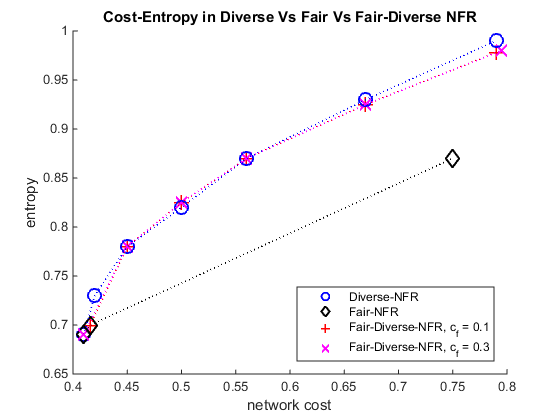}}
\caption{Normalized Network cost (x-axis) Vs Entropy (y-axis) in Diverse-NFR (blue), Fair-NFR (black), and Fair-Diverse-FR for (i) $c_f = 0.1$ (red), (ii) $c_f = 0.3$ (magenta) in the setup of \{Lastfm pop1 a0.8 N2 C20 Q0.8 L40\}.}
\label{fig3}
\end{figure}

Simulations were conducted for various fairness and diversity configurations by adjusting \( b \) and \( c_f \). The entropy vs. network cost trade-offs, comparing (i) Fair-NFR, (ii) Diverse-NFR, and (iii) Fair-Diverse-NFR (with \( F_{max} \) as the fairness constraint), are illustrated in Fig. \ref{fig3}.\\

\begin{table}[t]
\caption{Fairness metrics MAX, TV, and KL.}
\begin{center}
\begin{tabular}{c}
\hline
$\mathbf{F_{max}: } F_{max} = max_{i \in K} |p_i^{NF} - p_i^{BS}| \text{  (individual fairness)}$ \\
$\mathbf{F_{TV}: } F_{TV} = \frac{1}{2} \cdot \sum_{i \in K} |p_i^{NF} - p_i^{BS}| \text{  (total variation)}$ \\
$\mathbf{F_{KL}: } F_{KL} = \sum_{i \in K} p_i^{BS} \log(\frac{p_i^{BS}}{p_i^{NF}})\text{  (KL divergence)}$\\  
\hline
\end{tabular}
\label{tab6}
\end{center}
\end{table}

\begin{table}[t]
\caption{Linear fairness constraints $\mathbf{S}\left(\mathbf{z}, \mathbf{p}^{\mathrm{NF}}\right)$.}
\begin{center}
\begin{tabular}{c}
\hline
$\mathbf{F_{TV}}: \sum_{i \in K} z_i  \leq c_f,  p_i^{B S}-p_i^{N F} \leq z_i, p_i^{N F}-p_i^{B S} \leq z_i$ \\
$\mathbf{F_{KL}}: \sum_{i \in \mathcal{K}} p_i^{B S} \cdot z_i \geq-\left(c_f-\sum_{i \in \mathcal{K}} p_i^{B S} \log \left(p_i^{B S}\right)\right)$ \\
$z_i \leq e^{(m-1) \cdot s} \cdot p_i^{N F}-(m-1) s-1 $ \\
$\mathbf{F_{\max}}: p_i^{B S}-p_i^{N F} \leq c_f, p_i^{N F}-p_i^{B S} \leq c_f$\\  
\hline
\end{tabular}
\label{tab7}
\end{center}
\end{table}

\noindent The following key findings emerge from this analysis:\\

\noindent {\bf Key Finding 1: \emph{The diversity constraint serves a distinct purpose from fairness constraints in\cite{e17}.}} The Fair-NFR (black points) does not achieve the favorable cost-diversity trade-offs that our Diverse-NFR (blue points) achieves. This suggests that fairness metrics alone are insufficient for meeting diversity goals, and explicitly incorporating the entropy constraint is necessary.\\

\noindent {\bf Key Finding 2: \emph{Fairness and diversity constraints can coexist without sacrificing favorable trade-offs.}} Fig. \ref{fig3} shows that the favorable trade-off of Diverse-NFR (blue points) is maintained when applying fairness constraints—both strict (\( c_f = 0.1 \), red points) and less strict (\( c_f = 0.3 \), magenta points)—allowing for simultaneous fairness and diversity requirements to be effectively met.\\

{\bf Remark.}  The thresholds \( q \), \( b \), \( c_f \), and the fairness metric \( F \) can all be adjusted by the content provider according to its operational requirements. \\

\begin{figure}[t]
\centerline{\includegraphics[scale=0.6]{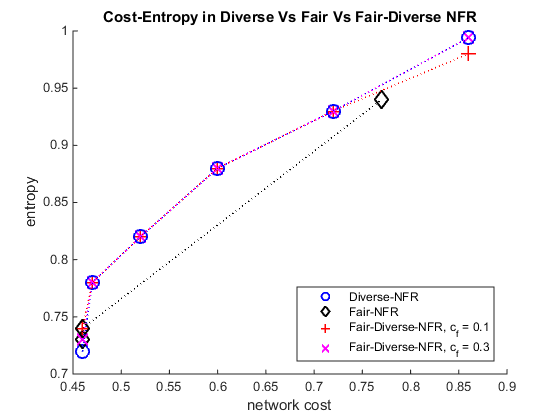}}
\caption{Normalized Network cost (x-axis) Vs Entropy (y-axis) in Diverse-NFR (blue), Fair-NFR (black), and Fair-Diverse-FR for (i) $c_f = 0.1$ (red), (ii) $c_f = 0.3$ (magenta) in the setup of \{Lastfm pop0 a0.8 N2 C20 Q0.8 L40\}}
\label{fig4}
\end{figure}

We note that the presented results are based on extensive simulations across all three fairness metrics (max, KL and TV), various values of \( b \), and a range of \( c_f \) constraints, covering both strict and lenient conditions. These values, along with the simulation parameters of Section III.A, do exert some influence on the outcomes. For example, a higher $pop$ value slightly improves the performance of the max-Fair-NFR. This is evident from the upward shift of the black line (i.e., max-Fair NFR without an entropy-based diversity constraint) in Figures \ref{fig3} and \ref{fig4} - which share \emph{all} simulation parameters except the value of \( pop \). However, in none of the results were the fairness constraints of the Fair-NFR able to match the performance of our Diverse-NFR alone, while they consistently facilitated diversity trade-offs when also included in the Diverse-NFR OP.\\

\section{CONCLUSION}
Previous studies have shown that NFR can greatly improve network performance, yet they often neglect the critical aspect of diversity, which is essential for users, content providers, and content creators alike. This work addresses this gap by incorporating diversity into NFR and demonstrating that a high level of diversity can be achieved with a lower increase in network cost. Moreover, we show that the proposed Diverse-NFR is compatible with existing fairness metrics from the literature\cite{e17}, enabling the simultaneous delivery of both diverse and fair recommendations.\\

We believe that the findings of this paper will spark further exploration of diversity within NFR. Future research could investigate: (i) diversity-cost trade-offs in different NFR schemes, e.g. NFR models that \emph{jointly} select recommendations and network policies, or scenarios where the likelihood of selecting a recommendation depends on its rank in the list; and (ii) alternative diversity metrics, such as the Gini index or the ``effective content pool'', i.e., the number of items with stationary access probabilities above a certain threshold, $p_i^{min}$.\\

\bibliography{references}

\vspace{12pt}

\end{document}